\documentclass[aps,amssymb,amsfonts,twocolumn,showpacs]{revtex4}
\usepackage{graphicx}

\newcommand{\ltsimeq}{\raisebox{-0.6ex}{$\,\stackrel
        {\raisebox{-.2ex}{$\textstyle <$}}{\sim}\,$}}
\newcommand{\gtsimeq}{\raisebox{-0.6ex}{$\,\stackrel
        {\raisebox{-.2ex}{$\textstyle >$}}{\sim}\,$}}

\begin{document}
\title{Collisions between solitary waves of three-dimensional Bose-Einstein condensates}
\author{N.G. Parker$^{1,2}$, A. M. Martin$^1$, S. L. Cornish$^2$ and C. S. Adams$^2$}
\affiliation{ $^1$ School of Physics, University of Melbourne,
Parkville, Victoria 3010, Australia \\ $^2$ Department of Physics,
University of Durham, South Road, Durham, DH1 3LE, United Kingdom }

\begin{abstract}
We study bright solitary waves of three
dimensional trapped Bose-Einstein condensates and their collisions. For a single solitary
wave, in addition to an upper critical number, we also find a {\em
lower} cut-off, below which no stable state can be found. Collisions
between solitary waves can be elastic, inelastic with either reduced
or increased outgoing speed, or completely unstable due to a collapse
instability. A $\pi$-phase difference between the waves
promotes elastic collisions, and gives excellent agreement with recent experimental
results over long timescales.
\end{abstract}


\maketitle
Bright solitary waves (BSWs) of atomic Bose-Einstein condensates (BECs) have been observed
in the form of a single BSW \cite{khaykovich} and groups of interacting BSWs
\cite{strecker,cornish_new} for attractive atomic interactions, and as gap BSWs in
periodically-confined BECs with repulsive interactions \cite{gap}.
In the former case, the attractive interactions which support the BSW against dispersion also induce a
collapse instability when the number of atoms exceeds a critical
value $N_{\rm C}$ \cite{bradley,roberts}, making attractive
BECs problematic to form. Consequently, `soliton'
experiments first form a stable BEC with repulsive atomic
interactions and switch to attractive interactions by means of the
molecular Feshbach resonance \cite{cornish}. For $N<N_{\rm C}$, a
single BSW forms \cite{khaykovich}, while for $N>N_{\rm C}$ a
`bosenova'-type collapse instability \cite{donley,roberts,gerton} is
induced with excess atoms being ejected from the condensate. The
`remnant' condensate is observed to contain BSWs
\cite{strecker,cornish_new} which oscillate in the trap and
interact with each other.  It has been inferred \cite{strecker} that
the BSWs repel each other due to $\pi$-phase differences between
them, thought to be formed by modulational instability of the
condensate \cite{strecker,khawaja_salasnich,carr}. With matter-wave `solitons' being potential candidates for applications
such as interferometry and quantum information processing, a thorough
understanding of their properties and in particular their {\em
interactions} is of key importance.

Theoretically, bright solitons are stable solutions of the 1D nonlinear
Schr\"{o}dinger equation with a sech-shaped wave profile
\cite{shabat}. In isolation their form is preserved by a balance
between dispersion and attractive nonlinearity, and they can have
any population. Another key property is that they emerge from
collisions with unchanged form. However, even in 1D, the exact
collisional form depends on the relative phase between the incoming
solitons $\Delta \phi$ \cite{gordon}: for $\Delta \phi=0$ the waves
can overlap completely (attractive solitonic interaction), while for $\Delta
\phi=\pi$ no overlap can occur and the solitons appear to `bounce'
(repulsive solitonic interaction). BSWs, the 3D analog of the bright
soliton, may have considerably different properties: the existence
of multiple dimensions leads to the collapse instability
\cite{bradley,donley} and thermal dissipation \cite{sinha}. Indeed,
two colliding BSWs can form a high-density state unstable to
collapse, with relative phase difference playing a key role
\cite{carr,salasnich}. Whereas the experiments of
\cite{strecker,khaykovich} use highly-elongated `quasi-1D' geometries,
Cornish {\it et al.}
\cite{cornish_new} employ a very weakly-elongated geometry, and, somewhat surprisingly,
observe well-behaved BSW oscillations for very long times with
negligible damping.  This raises questions over how `solitonic' such
3D states are.

In 3D an untrapped attractive BEC is always
unstable to collapse \cite{nozieres}.
Addition of confinement can stabilise it up to a critical
density and number of atoms $N_{\rm C}$
\cite{roberts,ruprecht,gammal,yukalov}. By definition a BSW should
be capable of axial self-trapping. In the absence of external axial trapping, radial confinement
is necessary to make $N_{\rm C}$ finite, and also to enable approximation to a 1D soliton.

In this work we simulate BSWs of 3D BECs and their collisions.
We observe rich and complex dynamics, which extend well
beyond the 1D soliton. Below a {\em lower} critical atom number we show that stable BSW solutions do not exist.
Collisions can be elastic, inelastic with increased or
decreased speed, or completely destructive. We
highlight the effect of relative phase, number of atoms and speed on
the collisional stability.  Furthermore we show that recent
observations of Cornish {\it et al.} \cite{cornish_new} are consistent with
two oscillating BSWs with $\pi$-phase difference, with the simulations
giving excellent agreement with the experimental oscillation data, over long timescales (for over 20 oscillations/3 seconds), despite neglecting
higher-order effects such as thermal dissipation \cite{sinha}.

We employ the Gross-Pitaevskii equation (GPE) to describe the
condensate mean-field `wavefunction' $\psi({\bf r},t)$,
\begin{equation}
i\hbar \frac{\partial \psi}{\partial
t}=\left[-\frac{\hbar^2}{2m}\nabla^2 + \frac{1}{2}m\left(\omega_r ^2
r^2 + \omega_z ^2 z^2\right)+g|\psi|^2 \right]\psi.
\end{equation}
Here $m$ is the atomic mass and $g=4\pi \hbar^2 a_{\rm S}/m$ is the
interaction coefficient, where $a_{\rm S}$ is the {\it
s}-wave scattering length ($a_{\rm S}<0$ in this study, corresponding to attractive atomic
interactions). The harmonic confining potential is
cylindrically-symmetric with $z$ and $r$ representing the axial and
radial coordinates. We base our analysis on the $^{85}{\rm Rb}$
experiment \cite{cornish_new} with
scattering length $a_{\rm S}=-0.6$~nm and radial trap frequency
$\omega_r=2\pi \times 17.5$~Hz.

\begin{figure}
\includegraphics[width=8.cm,clip]{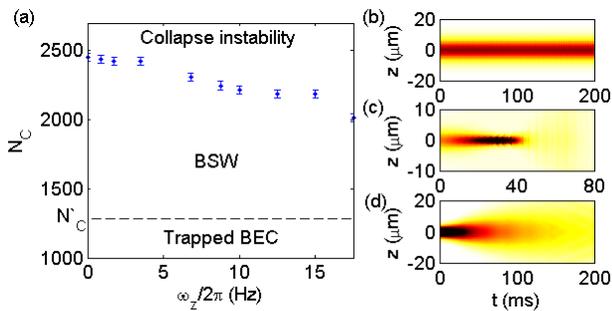}
\caption{(a) Upper critical number $N_{\rm C}$ as a function of
$\omega_z$ ($\omega_r=2\pi \times 17.5$~Hz). Indicated are the lower critical
number $N'_{\rm C}$ for $\omega_z=0$ (dashed line) and the condensate regimes. (b) Radially-integrated axial
density versus time for a BSW ($N_{\rm S}=1500$) with $\omega_z=0$.
(c)-(d) Same as (b) but with initial density
rescaled to (c) $N_{\rm S}=2700$ and (d) $N_{\rm S}=1000$. Dark/light regions represent high/low density.} \label{fig1}
\end{figure}

We begin by considering isolated BSW solutions. To generate such
states we propagate the GPE in imaginary time subject to the desired
atom number $N_{\rm S}$ and external potential \cite{imaginary
time}. Following previous studies \cite{ruprecht,gammal} we isolate
the critical number for collapse $N_{\rm C}$ when the GPE no longer
converges to a solution. Figure \ref{fig1}(a) shows the critical
number $N_{\rm C}$ as a function of axial trap frequency $\omega_z$.
For an axially-homogeneous system we find a range of values of
$N_{\rm S}$ where solutions exist which are axially self-trapping
(Fig.~\ref{fig1}(b)), i.e. BSW solutions. For $N_{\rm S}> N_{\rm
C}=(2450\pm 30)$ the wavepacket is unstable to interaction-induced
collapse, followed by explosive dynamics (Fig.~\ref{fig1}(c)). In
addition to the upper critical number we also find a lower
critical regime when $N_{\rm S}<N'_{\rm C}=(1280 \pm 20)$. For $N_{\rm S}<N'_{\rm C}$ the
wavepacket is no longer self-trapped axially but
spreads over time (Fig.~\ref{fig1}(d)), since the attractive
interactions are no longer sufficient to prevent dispersion of the wave. Both the low-$N_{\rm S}$ and high-$N_{\rm S}$ instabilities
do not occur for the 1D soliton, which is robust to any population.

Under axial confinement $N_{\rm C}$ decreases weakly
with $\omega_z$. The axial trapping increases
the peak density \cite{perez_garcia}, making the BSW more prone to the
collapse instability.  For $\omega_z=2\pi \times 6.8$~Hz
\cite{cornish_new} we find $N_{\rm
C}=(2305 \pm 30)$, in excellent agreement with experimental
observations of $N_{\rm C} \approx (2284 \pm 300)$ \cite{roberts}.
Alternative approaches have previously over or under estimated $N_{\rm C}$ \cite{gammal,yukalov}.

Axial confinement provides a sufficient restoring force to axially
trap the condensates and achieve stable solutions for $N<N'_{\rm
C}$. However, such a state can no longer be classed as a BSW because
it is not capable of axial self-trapping; it is then simply the
ground state trapped attractive BEC.  The regimes of the attractive
system, i.e. instability to collapse, stable BSW, and trapped BEC,
are indicated in Fig.~\ref{fig1}(a). The states observed in \cite{cornish_new} appear to have a population in the
range $N'_{\rm C}<N<N_{\rm C}$, consistent with being BSWs.

\begin{figure}
\includegraphics[width=8.cm,clip]{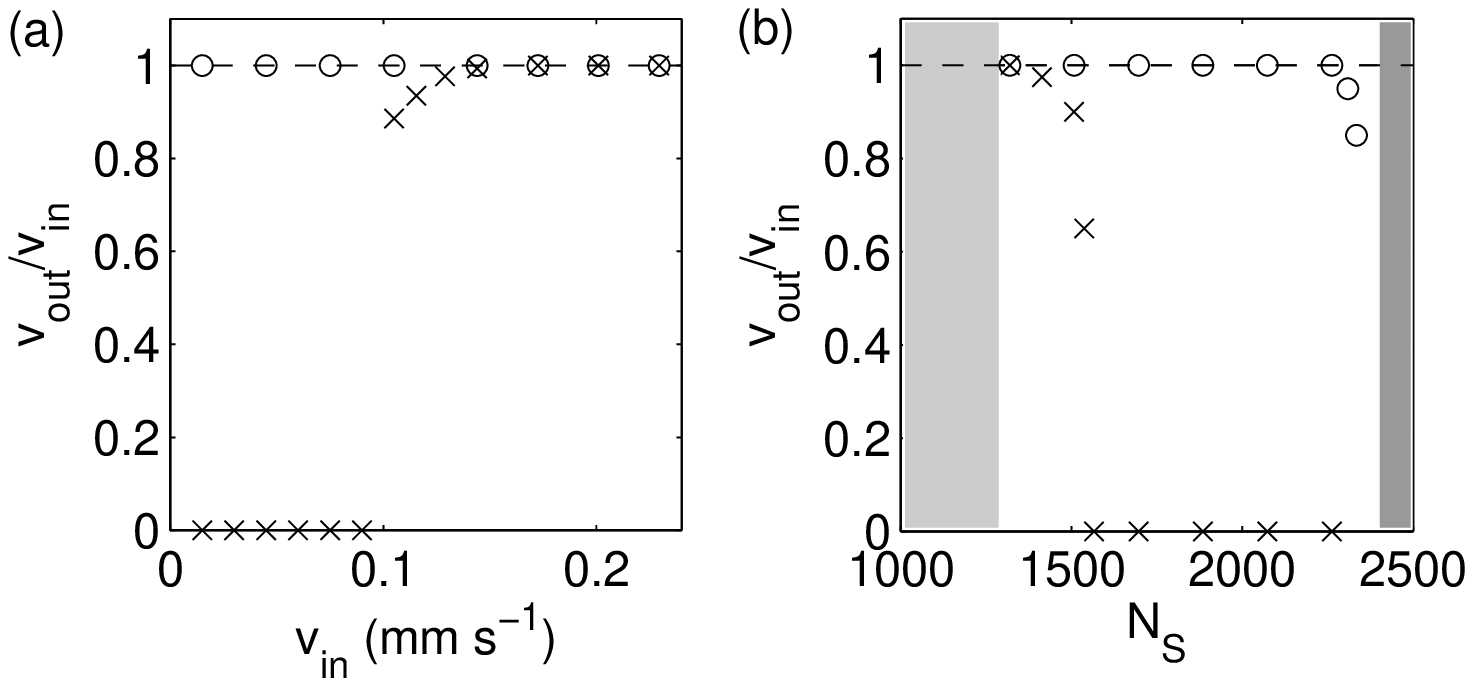}
\includegraphics[width=7.8cm,clip]{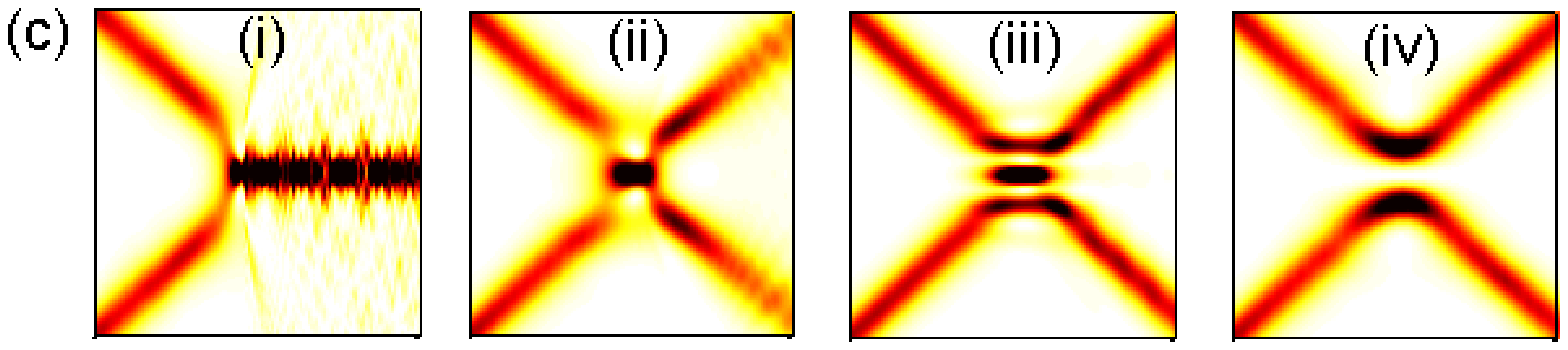}
\caption{(a)-(b) $v_{\rm out}/v_{\rm in}$ for a BSW collision
with $\Delta\phi=0$ (crosses) and $\Delta \phi=\pi$ (circles) in a
$\omega_z=0$ ($\omega_r=2\pi \times 17.5$~Hz) system as a function of (a) incoming speed
$v_{\rm in}$ ($N_{\rm S}=1500$), and (b) atom number $N_{\rm S}$
($v_{\rm in}=0.12~{\rm mm s^{-1}}$). The dashed line represents
$v_{\rm out}=v_{\rm in}$. In the shaded regions BSW solutions are
not accessible (light region: $N_{\rm S}<N'_{\rm C}$; dark region: $N_{\rm S}>N_{\rm C}$). (c) Axial density (integrated over $r$) as a
function of $v_{\rm in}t$ for (i) $\Delta \phi=0$ and
$v_{\rm in}=0.03{\rm mms}^{-1} $, (ii) $\Delta \phi=0$ and $v_{\rm
in}=0.09~{\rm mms}^{-1} $, (iii) $\Delta \phi=0$ and $v_{\rm
in}=0.24~{\rm mms}^{-1} $, and (iv) $\Delta \phi=\pi$ and $v_{\rm
in}=0.09~{\rm mms}^{-1} $. Each box represents the region $[0,80~\mu
{\rm m}/v_{\rm in}] \times [-40,40]~\mu{\rm m}$.} \label{fig2}
\end{figure}

We now consider collisions between two BSWs, each with $N_{\rm S}=1500$, in an axially-homogeneous system.
In 1D this collision is stable and elastic. Initially the BSWs are
well-separated and given an axial momentum kick,
$\psi=|\psi|\exp(imv_{\rm in}z/\hbar)$, such that they propagate
towards each other with speed $v_{\rm in}$.  In isolation the BSWs
propagate with constant speed and shape (see early times in Fig.
\ref{fig2}(c)). We also control the phase difference $\Delta
\phi$ between the BSWs by initially multiplying one BSW by $\exp(i\Delta \phi)$. We primarily
consider the extreme cases of $\Delta \phi=0$ and $\pi$. We consider
speeds of the order of those observed experimentally
\cite{cornish_new}: assuming an amplitude $z_0\sim 10~\mu$m in the trap,
harmonic motion suggests a maximum speed $v=\omega_z z_0\sim
0.4~{\rm mms}^{-1}$.

We parameterise the stability of a BSW collision in terms of the
outgoing speed $v_{\rm out}$. Figure \ref{fig2}(a) plots the ratio
$v_{\rm out}/v_{\rm in}$ as a function of incoming speed $v_{\rm
in}$ for the cases of $\Delta \phi=0$ (crosses) and $\Delta
\phi=\pi$ (circles). For $\Delta \phi=0$ and {\it low} $v_{\rm in}$ ($v_{\rm in}\ltsimeq 0.09~$mm s$^{-1}$), the BSWs do not emerge
from the collision (Fig.~\ref{fig2}(c)(i)) \cite{speed}. The overlap
of the BSWs forms an intermediate state with sufficiently high
density that a destructive bosenova-like collapse instability is induced
\cite{donley,carr,salasnich}. For
$\Delta \phi=0$ and {\it intermediate} $v_{\rm in}$ ($0.1 \ltsimeq v_{\rm
in} \ltsimeq 0.14~{\rm mm s^{-1}}$), BSWs emerge from the collision
but with $v_{\rm out}<v_{\rm in}$. The density again reaches a
critical level, but only a {\em partial} collapse instability is
induced (Fig.~\ref{fig2}(c)(ii)).  This leads to an irreversible transfer of energy from
centre-of-mass kinetic energy to excite 3D collective modes of the
BSWs. As the $v_{\rm in}$ is increased further the
collision becomes more elastic ($v_{\rm out}\rightarrow v_{\rm
in}$) and the shape excitations become negligible. These elastic
dynamics, shown in Fig.~\ref{fig2}(c)(iii), illustrate how the BSWs
pass through each other and form 3 fringes in a stable $\Delta \phi=0$ collision.
This limit is analogous to the corresponding 1D soliton collision.

In contrast, $\Delta \phi=\pi$ collisions are elastic throughout
this range of $v_{\rm in}$. Overlap of the BSWs at the centre of mass is
prevented and the BSWs `bounce' (Fig.~\ref{fig2}(c)(iv)). This
restricts the increase of the peak density during the collision
\cite{density}, and stabilises against instability
\cite{carr}.

From the solutions generated in Fig.~\ref{fig1}(a) for $\omega_z=0$ we estimate
the critical density for collapse to be $n_{\rm C}\sim 5\times 10^{19}~{\rm m^{-3}}$.
All of the $\Delta \phi=0$ collisions in Fig.~\ref{fig2}(a) exceed this density, while the $\Delta \phi=\pi$ collisions do not.  The stabilisation of collisions with speed suggests
that the {\em timescale} plays a key role. Indeed, in the elastic (inelastic) regime $v_{\rm in}\gtsimeq 0.15~{\rm mm s^{-1}}$ ($v_{\rm in}\ltsimeq 0.15~{\rm mm s^{-1}}$) the time over
which $n_{\rm C}$ is exceeded becomes consistently less (greater) than around $5~{\rm ms}$, which is the characteristic collapse time
observed experimentally \cite{donley}.

The stability is also sensitive to atom number $N_{\rm S}$. Figure
\ref{fig2}(b) shows $v_{\rm out}/v_{\rm in}$ as a function
of $N_{\rm S}$ ($v_{\rm in}=0.12~{\rm mm s^{-1}}$). Addition of atoms
makes the BSW more prone to the collapse instability,
destabilising the collisions with increasing $N_{\rm S}$. For $\Delta \phi=0$
the collisions break down for low $N_{\rm S}$ ($N_{\rm
S}\sim1400$) while the $\Delta \phi=\pi$ collision remains elastic
up until just below $N_{\rm C}$.

\begin{figure}
\includegraphics[width=8.0cm,clip]{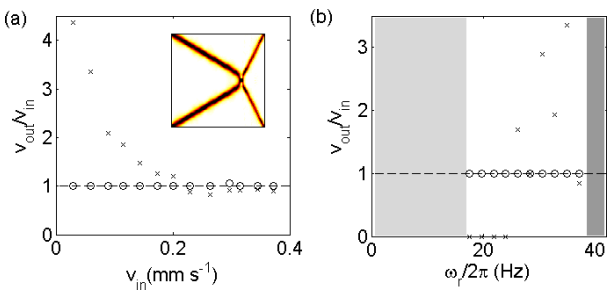}
\caption{(a)-(b) Ratio $v_{\rm out}/v_{\rm in}$ of a BSW collision with $\Delta\phi=0$ (crosses) and $\Delta \phi=\pi$
(circles) in a $\omega_z=0$ system as a function of (a) $v_{\rm in}$ (for $\omega_r=2\pi \times 35$~Hz and $N_{\rm S}=1500$) and (b)
$\omega_r$ (for $v_{\rm in}=0.06~{\rm mm s^{-1}}$ and $N_{\rm S}=1500$). The dashed lines plot $v_{\rm out}=v_{\rm in}$.
Inset in (a): axial density
(integrated over $r$) as a function of $v_{\rm in}t$ for $\Delta
\phi=0$, $v_{\rm in}=0.06~{\rm mm s^{-1}}$ and $\omega_r=2\pi \times
35$~Hz. The box represents the region $[-40,40]~\mu{\rm m} \times
[0,50~\mu{\rm m}/v_{\rm in}]$.  In the shaded regions of (b) BSW solutions are not accessible (light region: $N_{\rm S}<N'_{\rm C}$; dark region: $N_{\rm S}>N_{\rm C}$).
} \label{fig3}
\end{figure}

For $0<\Delta \phi<\pi$ the degree of overlap during the collision
varies between full overlap ($\Delta \phi=0$) and no overlap
($\Delta \phi =\pi$). Population transfer occurs between the BSWs,
with the outgoing speeds becoming asymmetric to conserve momentum, complicating the dynamics.
Although we concentrate here on $\Delta \phi=0$ and $\pi$ we have tested their robustness
by considering $10\%$ deviations of each ($\Delta
\phi=0.1\pi$ and $0.9\pi$), and observe that the regimes of
stability are robust to such variations \cite{pop}.

We now double the radial trap frequency to $\omega_r=2\pi \times
35$~Hz. The collisional stability as a function of incoming speed is
shown in Fig.~\ref{fig3}(a). For low $v_{\rm in}$ we again observe
inelastic collisions, but this time $v_{\rm out}> v_{\rm in}$ ($v_{\rm out}/v_{\rm in}\ltsimeq 4.5$). As the
incoming speed is increased further the collision becomes more
elastic ($v_{\rm out}\rightarrow v_{\rm in}$), as seen previously in
Fig.~\ref{fig2}(a).  A low speed inelastic collision with $v_{\rm
out}/v_{\rm in}\approx 3.3$ is shown in Fig.~\ref{fig3}(a)(inset).
During the collision the high density triggers a partial collapse
instability.  A small fraction of atoms (around $10\%$ in this case)
are ejected from the BSWs, releasing a significant amount of
interaction energy which is converted into kinetic energy. For lower $\omega_r$ this would tend to generate a radial
burst of highly-energetic atoms (bosenova) \cite{donley}.
However for sufficiently high $\omega_r$ we propose that this radial burst is suppressed, with a significant
fraction of this kinetic energy focussed axially into the BSWs, thereby
increasing their speed.
Again, for $\Delta \phi=\pi$ the collisions remain elastic
throughout this range of speeds.

In Fig.~\ref{fig3}(b) we vary the radial trap frequency $\omega_r$ while
keeping $N_{\rm S}$ fixed.  We observe BSW destruction for
$\Delta \phi=0$ and low $\omega_r$ (3D bosenova instability), while
for larger $\omega_r$ we typically see the ratio $v_{\rm out}/v_{\rm
in}$ grow. There is occasionally no speed increase,
highlighting the sensitivity of this highly nonlinear
effect. The $\Delta \phi=\pi$ collisions remain elastic.  For $\omega_r\sim 2\pi \times 38$~Hz, $N_{\rm C}$
decreases to $N_{\rm S}$ and we cannot generate BSW solutions beyond this.
If $N_{\rm C}$ is not exceeded all collisions
should become elastic in the quasi-1D limit of large $\omega_r$.

\begin{figure}
\includegraphics[width=8.5cm,clip=true]{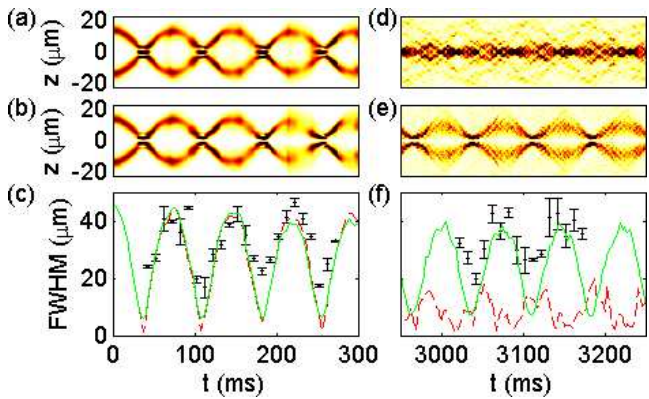}
\caption{(a)-(b) Evolution of the radially-integrated axial density
at early times for two BSWs ($N_{\rm S}=1500$ and initial amplitude $z_0=\pm 14\mu$m) with (a) $\Delta \phi=0$ and (b) $\Delta \phi=\pi$ ($\omega_z=2\pi \times 6.84$~Hz and $\omega_r=2\pi \times 17.5$~Hz).
(c) Full-width-half-maximum FWHM of a gaussian fit to the axial
density for $\Delta \phi=0$ (dashed line), $\Delta \phi=\pi$
(solid line), and experimental data (points)
\cite{cornish_new}. Note we have time-shifted the experimental data
to match the phase of the oscillations at early times. (d)-(f) Same
as (a)-(c) but for late times.} \label{fig4}
\end{figure}

In addition to radial trapping ($\omega_r=2\pi \times 17.5$~Hz) we now add axial confinement $\omega_z=2\pi \times 6.84$~Hz  to
compare with experiments \cite{cornish_new}.
Note we have slightly modified $\omega_z$ from the quoted experimental value of $\omega_z=2\pi \times 6.8$~Hz
in order to match the experimental data, as we will see.
It was observed that
the number of atoms left in the condensate remnant was greater than $N_{\rm C}$,
and divided between two states, each with population $N_{\rm
S}<N_{\rm C}$, which oscillate axially. We continue to employ $N_{\rm S}=1500$, which satisfies these conditions.
Note that the collapse-induced conversion of interaction into kinetic
energy discussed above may explain the formation of
oscillating BSWs in the experiment rather than stationary ones.
The BSWs are positioned at $z_0=\pm 14~\mu$m. We no longer
apply a momentum kick since the trap accelerates the BSWs. The
dynamics at {\em early times} for the cases of $\Delta\phi=0$ and
$\Delta \phi=\pi$ are shown in Fig.~\ref{fig4}(a) and (b)
respectively. Initially the BSWs accelerate identically towards the
trap centre. As they interact we see the different collisional
dynamics, with the BSWs overlapping for $\Delta \phi=0$ and `bouncing'
for $\Delta \phi=\pi$. However, at these early times the overall
dynamics for both cases are similar: the BSWs emerge from the
collisions with large collective excitations being generated. In
Fig.~\ref{fig4}(c) we plot the early evolution of the axial full width half maximum FWHM
(related to a gaussian fit) of the $\Delta \phi=0$ (dashed line) and $\Delta \phi=\pi$ (solid line) systems and compare
with the experimental data \cite{cornish_new} (circles). At
early times there is little difference between the $\Delta \phi=0$
and $\pi$ dynamics.  The dominant feature in both the experimental and numerical data is an
oscillation at $2\omega_z$, corresponding to the centre-of-mass
oscillations of the BSWs. We have chosen the initial BSW separation to approximately match the maximum experimental widths. The experimental widths
decrease to around $18 \mu {\rm m}$ which represents the resolution limit
of the measurements. The numerical minima occur at ${\rm FWHM}\sim
2-5~\mu{\rm m}$ (this is typically larger for $\Delta \phi=\pi$ due to the
repulsive nature of the interaction).  Whereas our simulations begin
at maximum widths, the experimental data begins at a minimum
\cite{cornish_new}, indicating that the BSWs are created in close
proximity. Note that in order to match the phase of the oscillations
in Fig.~\ref{fig4}(c) we have shifted the experimental data in time.

We do not detect the secondary $2\omega_r$ frequency component observed experimentally.
However, atoms ejected during
the initial experimental collapse lead to a significant oscillation at
$2\omega_r$ \cite{cornish_new}. Furthermore we have verified that any offset of the BSWs from the {\it
z-}axis will induce a $2\omega_r$ oscillation. The second largest frequency component in
the simulations occurs at $\omega\approx 2\pi \times 27$~Hz with an
amplitude of only around $5\%$ of the $2\omega_z$ component, and
arises from the collective modes excited in the BSWs.

Importantly, the experiment oscillations were observed for over 3~s with no discernable damping \cite{cornish_new}.
The dynamics at {\em late} times are shown in
Fig.~\ref{fig4}(d)-(f). The repeated collisions gradually destabilise
the collisions through the growth of collective excitations.  Partial
collapse instabilities eject highly energetic atoms from the BSWs, which
leads to the formation of a diffuse spread of low density `hot' atoms.
The $\Delta \phi=0$ case
is most prone to instability, and by late times the BSWs have
collapsed to a central highly-unstable wavepacket (Fig.~\ref{fig4}(c)). The FWHM for this
case lies well below the experimental values (Fig.~\ref{fig4}(f), dashed line). For $\Delta \phi=\pi$, two oscillating wavepackets
persist at late times, albeit in highly-excited states (Fig.~\ref{fig4}(e)).
The FWHM for $\Delta \phi=\pi$
(Fig.~\ref{fig4}(f), solid line) remains close to the
experimental values even at these late times.  Note that in order to match the phase of
the experimental data at late times we find it necessary to modify the quoted experimental axial trap
frequency by around $0.5\%$ to $\omega_z=2\pi \times 6.84$~Hz.  This
agrees well with the {\em fitted} frequency in the experiment
\cite{cornish_new}, and implies that the BSW oscillations are an accurate measurement of trap oscillation frequency.

In conclusion we find that bright solitary waves exhibit rich and complex
behaviour, not present for 1D solitons. We observe not just the upper
critical population familiar with trapped attractive BECs but also a {\em lower}
critical population, below which the wavepacket is no longer capable of axially self-trapping.
High-density collisions between BSWs induce instabilities. For large instability the BSWs
are destroyed by a catastrophic collapse. For intermediate instability the collisions are inelastic,
with {\em higher} or {\em lower} outgoing speed.  We highlight the sensitive dependence of the
instability on the phase difference between the BSWs, collisional speed, atomic population, and trap geometry, all of which are controllable experimentally.
In particular elastic collisions are promoted for a $\pi$-phase difference between the BSWs and high collisional speeds.
Furthermore, we find that the experimental observations of long-lived `soliton' oscillations by Cornish {\it et al.}
\cite{cornish_new} are successfully modelled in terms of two oscillating BSWs with a $\pi$-phase difference.

We acknowledge the UK EPSRC (NGP/CSA), Royal Society (SLC),
University of Melbourne (NGP/AMM) and ARC (NGP/AMM) for support. We
thank S. A. Gardiner and J. Brand for discussions.


\begin{thebibliography}{99}
\bibitem{imaginary time} Using an appropriate initial guess, propagation of the GPE in imaginary
time will converge to the ground state of the system, provided it
exists.
\bibitem{speed} When BSWs do not emerge from a collision we assign $v_{\rm out}=0$.
\bibitem{pop} Population transfer and speed change are typically of the order of
$1\%$.
\end{thebibliography}
\end{document}